\documentclass[lettersize,journal]{IEEEtran}
\makeatletter
\usepackage{amsmath}
\renewcommand{\maketag@@@}[1]{\hbox{\m@th\normalsize\normalfont#1}}%
\makeatother

\usepackage[utf8]{inputenc}

\usepackage{amsmath,amsfonts}
\usepackage{algorithmic}
\usepackage{algorithm}
\usepackage{array}
\usepackage[caption=false,font=normalsize,labelfont=sf,textfont=sf]{subfig}
\usepackage{textcomp}
\usepackage{stfloats}
\usepackage{xcolor}
\usepackage{url}
\usepackage{verbatim}
\usepackage{indentfirst}
\usepackage{graphicx}
\usepackage{cite}
\usepackage{pifont}
\usepackage{booktabs}
\usepackage{titlesec} 
\usepackage{hyperref}

\renewcommand\thesection{\arabic{section}}
\renewcommand\thesubsection{\thesection.\arabic{subsection}}
\renewcommand\thesubsubsection{\thesubsection.\arabic{subsubsection}}

\titleformat{\section}{\large\bfseries}{\thesection}{1em}{}
\titleformat{\subsection}{\normalsize\bfseries}{\thesubsection}{1em}{}
\titleformat{\subsubsection}{\normalsize\itshape}{\thesubsubsection}{1em}{}

\begin{document}

\title{A Secure Communication Protocol for Remote Keyless Entry System with Adaptive Adjustment of Transmission Parameters}

\author{
    Jingjing Guo, Bo Tang, Jiayuan Xu, Qingyi Li, Yuyuan Qin, Xinghua Li \\
    School of Cyber Engineering, Xidian University, Xi'an, China, 710000
}

\markboth{}%
{Shell \MakeLowercase{\textit{et al.}}: A Sample Article Using IEEEtran.cls for IEEE Journals}


\maketitle

\begin{abstract}
Remote Keyless Entry (RKE) systems have become a standard feature in modern vehicles, yet their unidirectional fixed-frequency radio communication renders them vulnerable to replay attacks, impersonation attacks, cryptanalysis, and intentional interference. Existing cryptographic authentication methods enhance security but often fail to address real-world constraints such as computational efficiency and radio interference. To mitigate these threats, we designed the Adaptive Frequency-Hopping Algorithm and the Adaptive TXP and PHY Mode Control Algorithm that can dynamically optimize channel selection, transmission power, and PHY modes based on real-time channel quality assessment. To enhance the security and reliability of RKE systems, we propose the Lightweight Vehicle-Key Authentication Protocol. In addition, a prototype of the proposed scheme was implemented to verify its effectiveness in mitigating interference and preventing unauthorized access.Experimental results show that our scheme significantly enhances communication security and reliability while maintaining low computational overhead. Under mild interference conditions, the packet delivery rate (PDR) of the adaptive scheme increases from 93\% to 99.23\%, and under strong interference, it improves from 85\% to 99.01\%. Additionally, the scheme effectively prevents replay and impersonation attacks, ensuring secure vehicle access control by dynamically optimizing communication parameters to maintain stable and reliable transmission.

\end{abstract}

\section{Introduction}
\subsection{Background}
Car Remote Keyless Entry (RKE) systems, which are standard or optional in 80\% of cars today\cite{ref1}, typically rely on short-range radio transmission using Bluetooth Low Energy (BLE) technology. These systems include a mobile, handheld car key fob with a short-range radio transmitter and a receiver inside the vehicle. RKE systems in North American cars typically operate at 315 MHz, while European and Asian cars use 433 MHz and 868 MHz, respectively. Using BLE technology, the key fob sends a radio frequency signal to the vehicle’s receiver as far as 80 meters away, allowing the motorist to unlock and lock the car doors without using a key. Upon pressing the key fob button, RF signals in the UHF frequency band, including pre-guidance codes, command codes, and encrypted scrolling codes, are sent. The vehicle’s antenna receives the signals, which are decoded by the body control module. The module verifies data’s validity, and the actuator controls the lock mechanism. Despite its convenience, this system’s reliance on BLE technology exposes it to various security threats and potential interference\cite{4}.

\subsection{Motivation}Remote Keyless Entry (RKE) systems have evolved to improve vehicle security, yet they remain vulnerable to sophisticated attacks. Early RKE systems relied on fixed codes, making them highly susceptible to replay attacks using software-defined radio (SDR) tools like HackRF\cite{5}. To address this, rolling-code mechanisms were introduced, generating unique one-time codes based on a counter synchronization system.

Despite this improvement, rolling-code RKE systems still face significant security threats. Attacks such as RollJam\cite{6} can intercept and replay valid codes, effectively bypassing rolling-code protections. Similarly, cryptanalytic techniques targeting NXP’s Hitag-2 transponders expose weaknesses in widely used vehicle models. The Rolling-PWN vulnerability further demonstrates that attackers can exploit synchronization flaws to gain unauthorized access.

Beyond cryptographic vulnerabilities, RKE systems are also susceptible to radio jamming, which can prevent vehicles from locking, increasing the risk of theft. Additionally, their reliance on fixed transmission frequencies makes signal interception and manipulation easier.

A key limitation of current RKE systems is their unidirectional communication model, which lacks mutual authentication. This allows attackers to forge commands without verification. Given these persistent risks, there is a clear need for enhanced RKE security mechanisms, incorporating bidirectional authentication, dynamic frequency adaptation, and stronger cryptographic defenses to counter emerging threats.

\subsection{Contributions}
In order to improve the security and communication stability of the RKE system, we propose a secure communication protocol for remote keyless entry systems
with adaptive adjustment of transmission parameters. Specifically, our contribution is reflected in the following three aspects:

(1) In the environment of Bluetooth frequency-hopping communication, after several experiments, we obtained a close relationship between the important parameters involved. As a result, we have designed an adaptive communication scheme suitable for RKE systems. This scheme adaptively selects good channels and selects physical layer modes and transmission powers by real-time evaluation of wireless channel quality.

(2) From the perspective of the limited computing and storage resources of RKE devices, we designed a lightweight Vehicle-Key authentication protocol based on ECDSA and AES. This protocol ensures the security of communication between the vehicle and the key, and can effectively prevent replay attacks and impersonation attacks.

(3) Based on the proposed adaptive RKE secure communication scheme, an adaptive RKE secure communication prototype system was implemented. The system includes two modules: adaptive communication and authentication. The prototype system was tested from three aspects: performance testing, overhead testing, and security analysis.

\section{Related Work}
Existing research on RKE system security has made significant progress in mitigating threats such as replay attacks. Various cryptographic authentication methods, including hash functions, asymmetric encryption, and symmetric encryption, have been proposed to enhance the security of RKE systems. However, these approaches often fail to address interference issues effectively. Additionally, some solutions require complex system deployment and expensive hardware, making practical implementation challenging.

Cryptographic-Based Authentication Mechanisms have been widely explored to protect RKE systems from replay and RollJam attacks. Several studies have proposed authentication schemes based on cryptographic techniques. For instance, Reference \cite{1} employs a hash-based and asymmetric encryption authentication mechanism, while Reference \cite{ref2} introduces a lightweight symmetric encryption protocol to defend against scanning attacks, replay attacks, and dual-thief attacks. Another study \cite{ref3} leverages quantum key distribution to develop a quantum-secure authentication protocol, which not only prevents replay and RollJam attacks but also offers protection against future quantum computing threats. However, while these cryptographic approaches enhance security, they do not address interference issues in RKE communication. Moreover, the quantum cryptographic system requires highly complex and expensive equipment, making it impractical for real-world deployment.

Physical Unclonable Function (PUF)-Based Approaches have also been proposed as a security mechanism for RKE systems. Reference \cite{ref4} introduces a PUF-based RKE system that enhances security by ensuring the privacy of communication entities. Similarly, Reference \cite{ref5} presents a PUF-based authentication protocol capable of preventing replay, RollJam, and RollBack attacks. However, despite their advantages, PUF-based solutions still fail to address interference issues. Additionally, the inherent instability and noise characteristics of PUFs may affect the reliability and consistency of the generated keys, limiting their effectiveness in practical applications.

Improvements to Existing RKE Systems have been suggested to enhance both security and interference resistance. Reference \cite{ref6} applies the 2ASK optimal receiver theory to insert blank characters into RKE/PKE transmissions, thereby improving resistance to interference. This approach, combined with AES encryption, enhances security levels. Reference \cite{ref7} proposes a rolling code and timestamp-based solution, where the rolling code, command, and clock time are concatenated and encrypted using AES to counter replay attacks. Another study \cite{ref8} designs an enhanced RKE system using timestamps and XOR encoding to resist both replay and interference-replay attacks. While these approaches improve anti-interference capabilities to some extent, they do not completely eliminate interference issues in RKE communication.

Alternative Approaches beyond traditional cryptographic and PUF-based methods have also been explored. Reference \cite{ref9} proposes a blockchain-based approach to counter replay attacks, leveraging the decentralized nature of blockchain for security. However, blockchain networks often suffer from throughput and latency limitations, affecting their performance in real-time applications. Reference \cite{ref10} develops a digital car key system incorporating hardware-based security modules, protecting against replay, man-in-the-middle, and tampering attacks. However, this approach relies on electronic devices, meaning users may lose access to their vehicles if the device is lost, damaged, or runs out of battery. Additionally, digital keys remain vulnerable to hacking attempts. Reference \cite{ref11} introduces a physical-layer security method that utilizes directional antennas to generate interference and reduce the eavesdropper's signal quality, thereby preventing the successful decoding of data packets. Despite its effectiveness, this solution requires the installation of additional directional antennas, increasing system complexity and deployment costs.

In summary, while existing research has contributed to improving the security of RKE systems, significant challenges remain, particularly in mitigating interference attacks, cryptanalysis threats, and tampering risks. Additionally, many solutions require costly hardware and complex system architectures, limiting their feasibility in practical implementations.
\section{System Framework}\label{333}

\subsection{System model}
\begin{figure*}[h]
    \centering
    \includegraphics[width=1\textwidth]{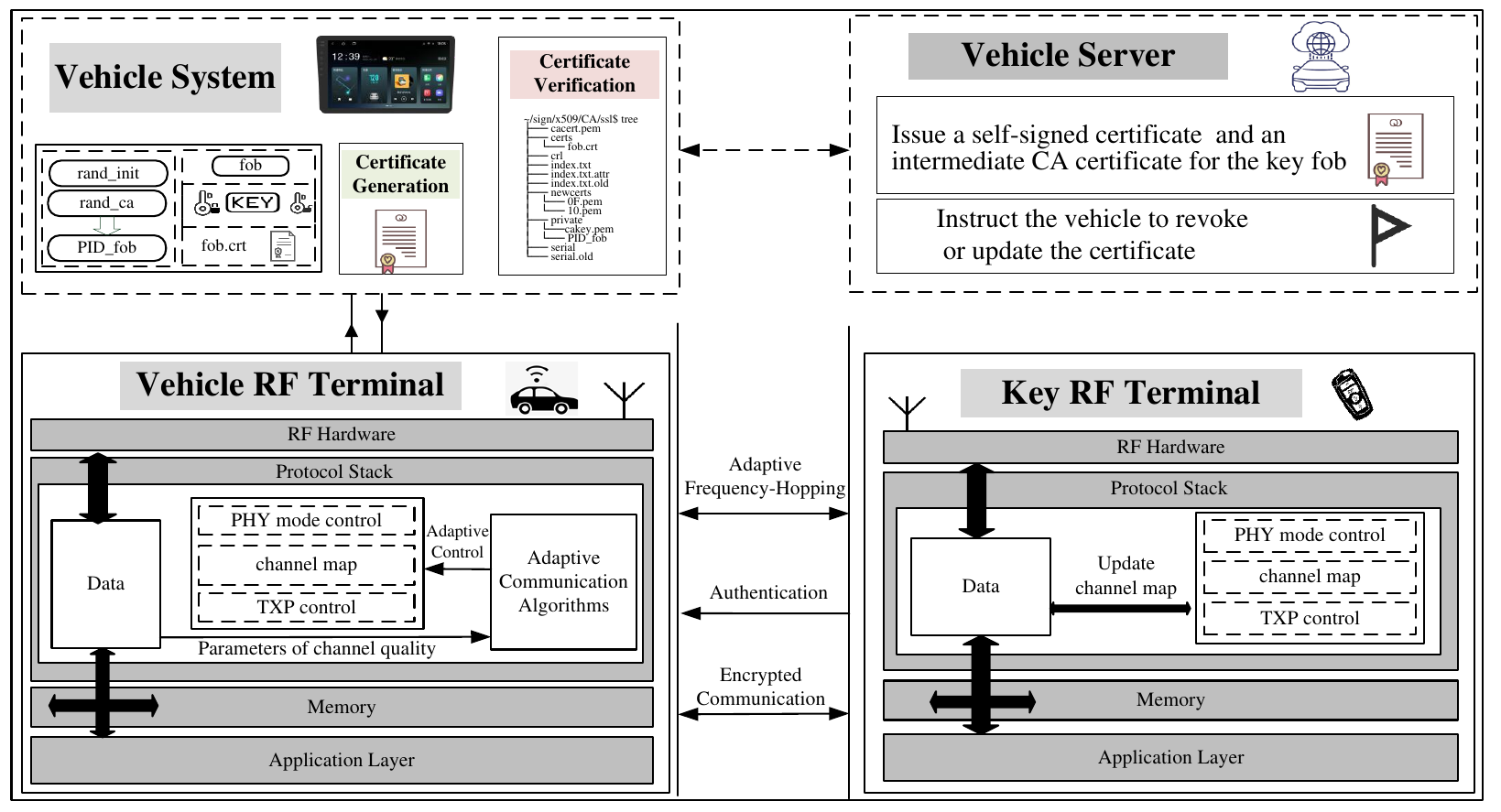}
    \caption{System Work Model}
    \label{fig:flowchart}
\end{figure*}

This system focuses on secure communication and certificate management for vehicle key systems. Figure \ref{fig:flowchart} consists of four components: the Vehicle Server (Root CA), the Vehicle System (Intermediate CA), the Vehicle RF Terminal, and the Key RF Terminal (key fob). The functions of each part are described in detail below. Symbols used in the text are explained in the Table \ref{1}.

\begin{table}[htbp]
\hspace{-5mm}
\centering
\caption{Symbol Table}
\begin{tabular}{|c|c|}
\toprule
\textbf{Symbol} & \textbf{Meaning} \\
\midrule
$fob.pk$ & the public key of key fob\\
$fob.sk$ & the private key of key fob \\
$PIDfob$ & Pseudo ID of the car key \\
$rand\_{ca}$ & Random number used to generate $PIDfob$ \\
$rand\_{init}$ & AES encryption/decryption key \\
$Enc\_{AES}[rand\_{init}, *]$ & AES encryption with the key $rand\_init$ \\
$Dec\_{AES}[rand\_{init}, *]$ & AES decryption with the key $rand\_init$ \\
$* || *$ & Concatenate two messages \\
$\oplus$ & XOR (Exclusive OR) \\
$Gen(rand)$ & Generate random number $rand$ \\
$Verify(a, b)$ & Verify if $a$ and $b$ are equal \\
$Verify(fob.crt)$ & Verify car key certificate \\
\bottomrule
\end{tabular}
\label{1}
\end{table}

\paragraph{Vehicle Server}
The vehicle server acts as the root certificate authority (root CA), managing certificates and issuing them to the vehicle system. Therefore, It acts as the root CA for issuing certificates to the vehicle's system (intermediate CA) and instructing the vehicle system to revoke or update the key's certificate.

\paragraph{Vehicle System}
The vehicle system acts as an intermediate certificate authority (intermediate CA). After the vehicle server issues the intermediate CA certificate, it then issues a certificate for the key fob.Therefore, the vehicle system performs four main functions:
\begin{itemize}
    \item Acting as an intermediate CA to issue, update, or revoke the key certificate as instructed by the server.
    \item Generating the secret key for initial communication between vehicle RF Terminal and Key RF Terminal.
    \item Constructing the certificate authentication system to verify the key's legitimacy.
\end{itemize}

\paragraph{Vehicle RF Terminal}
The vehicle RF terminal verifies communication keys, establishes secure connections, adapts communication, authenticates with the vehicle system. Therefore, the vehicle RF terminal performs two main roles: 

\begin{itemize}
    \item Scanning the broadcast to establish a secure connection and assist in key authentication with the RF terminal.
    \item Generating adaptive communication parameters and encrypting communication with the key RF terminal.

\end{itemize}

\paragraph{Key RF Terminal}
The key RF terminal broadcasts to connect with the vehicle RF terminal, then uses adaptive frequency-hopping for secure communication. It sends its certificate for authentication and, once verified, transmits encrypted commands. During updates, it securely receives and updates key data from the vehicle system.The key RF terminal performs four main functions: 
\begin{itemize}
    \item Broadcasting to establish a secure connection and send authentication parameters for verification.
    \item Performing adaptive and encrypted communication with the vehicle RF terminal.
\end{itemize}

\subsection{Design Goal} \label{3.2}
The proposed RKE system aims to achieve three key objectives: 

(1) Robust communication through real-time channel adaptation between RF terminals. 

(2) Strong security via certificate-based authentication and AES encryption managed by the vehicle server and system.

(3) Efficient operation optimized for resource-constrained key fobs. 

These goals guide the system's workflow from initialization to command transmission and certificate updates.

\subsection{Work Flow}\label{3.333}

Based on the model described above, this section introduces the working process of the proposed RKE secure communication system. Figure \ref{fig:System work flow chart} presents the workflow of the proposed adaptive RKE secure communication system. In the proposed scheme, the RKE system consists of four components: the vehicle server, the vehicle system, the vehicle RF terminal, and the key fob. According to the system execution process, there are a total of six phases: 

\begin{figure*}[h]
    \centering
    \includegraphics[width=0.7\textwidth]{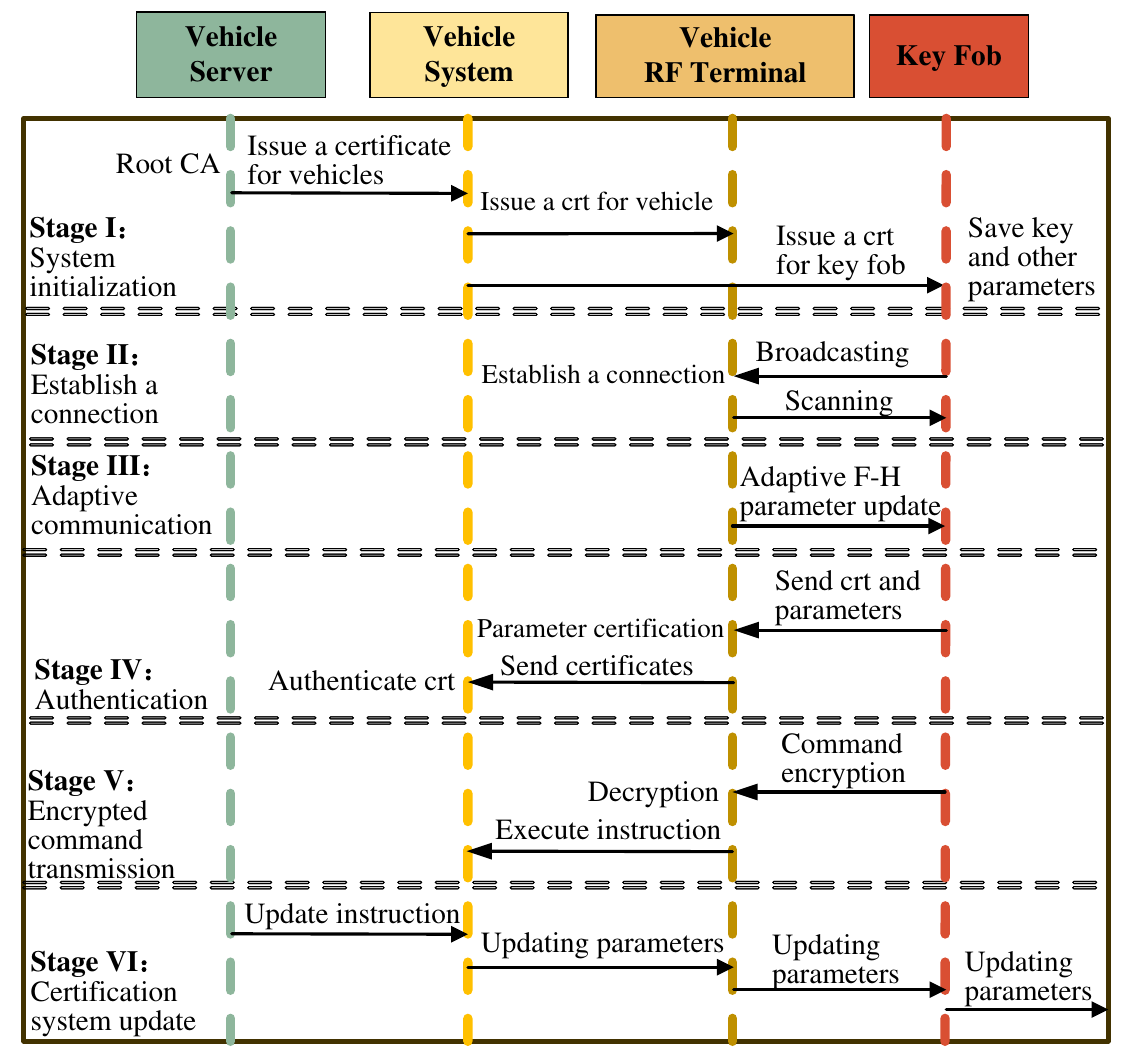}
    \caption{System work flow chart}
    \label{fig:System work flow chart}
\end{figure*}

     \textbf{Step 1: System Initialization}
    
In the initialization phase of the adaptive frequency-hopping RKE security system, the initialization of four systems—vehicle server, vehicle system, vehicle RF terminal, and key fob—is performed. The vehicle server acts as the root CA (Certificate Authority), generating a public-private key pair for the root CA $(ca.pk, ca.sk)$ and issuing a self-signed certificate for itself $(ca.crt)$. The vehicle system acts as an intermediate CA, generating an intermediate CA private key $(car.sk)$ and certificate request $(car.csr)$, which is signed by the vehicle server to issue an intermediate CA certificate $(car.crt)$.

After receiving the intermediate CA certificate $(car.crt)$ from the vehicle server, the vehicle system first generates random numbers $(rand\_ca)$ and $(rand\_init)$, then generates the key fob’s pseudo-ID $(PIDfob)$ and public-private key pair $(fob.sk, fob.pk)$. The $(rand\_ca)$ is used to generate $(PIDfob)$. The key certificate request file $(fob.csr)$ is generated based on the pseudo-ID of the key and its public-private key. Finally, the vehicle system signs the key fob certificate $(fob.crt)$ with the intermediate CA certificate $(car.crt)$. The generated certificate files are saved to the vehicle system’s certificate management system, completing the initialization of the vehicle system’s authentication system. These certificates and key files $(rand\_init, PIDfob, fob.sk, fob.pk)$ are transmitted to the vehicle RF terminal and key fob.

\begin{equation}
PIDfob = IDfob \oplus VIN \oplus rand\_ca 
\label{111}
\end{equation}

The vehicle RF terminal determines the adaptive communication parameters and deploys the adaptive communication system. The $(PIDfob)$ broadcast address is added to the whitelist, and $(PIDfob)$ and $(rand\_init)$ are saved to the key storage area, completing the initialization of the vehicle RF terminal.

The key fob saves $(rand\_init)$, the public-private key pair $(fob.sk, fob.pk)$, the pseudo-ID $(PIDfob)$, and the key fob certificate $(fob.crt)$ to the key storage area, completing the key fob initialization.

    \textbf{Step 2: Connection Establishment} 

During the connection establishment phase, the key fob uses the AES algorithm to encrypt $(PIDfob)$ and broadcast it, with $(rand\_init)$ as the encryption key. The vehicle RF terminal scans for broadcasts and uses the saved initial symmetric key $(rand\_init)$ to decrypt and compare $(PIDfob)$. Once successful, a secure connection is established between the vehicle RF terminal and the key fob.

    \textbf{Step 3: Adaptive Communication}
    
In the adaptive frequency-hopping channel selection phase, the vehicle RF terminal monitors channel quality, generates communication system parameters such as the frequency-hopping communication map (channel map), power, PHY, etc., and sends them to the key fob. Adaptive frequency-hopping communication is performed according to the predefined parameters.

    \textbf{Step 4: Identity Authentication}\label{5} 

The key fob and vehicle RF terminal exchange authentication information (such as random number verification), after which the key fob sends its certificate to the vehicle RF terminal, which then sends the certificate to the vehicle system to assist with the authentication process.

    \textbf{Step 5: Encrypted Transmission of Control Commands}\label{5} 

After successful authentication, the key fob uses the AES algorithm to encrypt the random number and control commands. The vehicle RF terminal receives the message, decrypts the control commands and random number, and verifies the random number before executing the control commands.

    \textbf{Step 6: Authentication System Update} 

When the vehicle server detects that the certificate needs to be updated, it notifies the vehicle system to update the certificate. The vehicle system generates new key-related parameters and certificates. After the vehicle and key fob are connected, the vehicle system sends the new key-related parameters and certificates to the vehicle RF terminal, which encrypts the data and sends it to the key fob. The key fob accepts the message, decrypts it, and updates the parameters and certificates.

\section{Our Work}
In Section \ref{333}, we describe the system framework and workflow in detail. Steps 4 and 5 in Section \ref{3.333} of the workflow are our main tasks. In the following Section \ref{555}, we explain in detail the work of the adaptive communication part in step 4, and in Section \ref{666} we explain in detail the lightweight Vehicle-Key authentication protocol part in step 5.

\section{Adaptive Communication Scheme} \label{555}

This section presents a comprehensive adaptive communication framework designed to enhance the robustness and efficiency of BLE-based systems.The following subsections detail the fundamental methodologies, including channel quality data collection mechanisms, parameter correlation analysis, and adaptive algorithm designs.

\subsection{BLE Channel Quality Data Collection and Parameter Analysis} 

Effective channel adaptation relies on systematic data collection and rigorous parameter analysis.  We first introduce the methodology for acquiring channel quality indicators, followed by an in-depth examination of critical parameters influencing BLE communication performance. The analysis provides essential insights for developing the subsequent adaptive algorithms.
\subsubsection{Data Collection}

Channel quality data collection is the foundation for achieving channel adaptation, PHY mode adaptation, and transmission power adaptation. The effectiveness and reliability of adaptive algorithms depend on accurate channel quality data. By continuously collecting parameters such as signal strength, signal-to-noise ratio (SNR), and bit error rate (BER), the system can effectively adjust frequency-hopping channels, PHY modes, and transmission power to maximize communication performance. The accuracy and timeliness of channel quality data directly impact the system's response speed and effectiveness in adapting to environmental changes. Therefore, channel quality data collection is a crucial step to ensure the efficient operation of adaptive algorithms.

Spörk et al.\cite{ref40} demonstrated that the packet delivery rate (PDR) at the link layer is an accurate estimator of Bluetooth Low Energy (BLE) channel quality. The PDR of each channel represents the transmission quality over a certain period, making the collection of PDR for each channel the first step in implementing adaptive frequency-hopping.

According to Section 2.4, the PDR of each channel can be calculated using Equation~\eqref{eq:pdr}:

\begin{equation}
PDR = \frac{\#ACK(P \to C)}{\#TX(C \to P)}
\label{eq:pdr}
\end{equation}

\begin{figure}[h]
    \centering
    \includegraphics[width=0.48\textwidth]{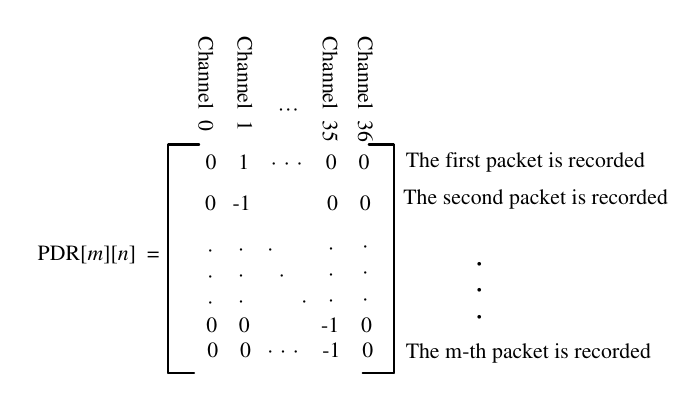}
    \caption{Channel transmission record quality matrix}
    \label{fig:flowchart1}
\end{figure}

To estimate the transmission quality of each channel in real time, we use a matrix $PDR\_latest\_array[m][n]$ to record the transmission status of $m$ packets on each channel, as shown in Figure \ref{fig:flowchart1}. The columns $n$ of the matrix $PDR\_latest[m][n]$ correspond to Bluetooth channels 0 to 36, while rows from 1 to $m$ represent the recorded first to $m$-th packets. 

After data collection, we count the number of ``1'' in each channel, representing the number of successfully transmitted packets $n\_ok$, and the number of ``-1'', representing the number of failed packets $n\_failed$. The real-time PDR of channel $i$ is then computed using Equation~\eqref{eq:pdr_latest}:

\begin{equation}
PDR\_latest[i] = \frac{n\_ok(i)}{n\_fail(i) + n\_ok(i)} \times 100\%
\label{eq:pdr_latest}
\end{equation}

\begin{figure}[h]
    \centering
    \includegraphics[width=0.48\textwidth]{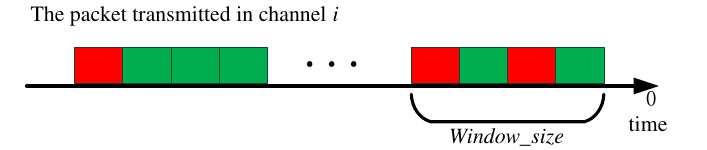}
    \caption{Schematic of the window length of the data record}
    \label{fig:windowsize}
\end{figure}

Matrix $PDR\_latest\_array[m][n]$ “m” is also called the “window length” ($window\_size$) of the data record. $window\_size$ represents the transmission of the past $window\_size$ data packets on this channel. As shown in Figure \ref{fig:windowsize}, $window\_size$ is set to 4, which means that 4 packets have been recorded on this channel. The transmission of the first and third packets failed, while the second and fourth packets were successfully transmitted. The smaller the window size, the more real-time the transmission, but the less accurate the channel accuracy estimate. In the channel quality collection matrix $PDR\_latest\_array[m][n]$, the larger the value of $window\_size$, the more accurate the calculation of the channel quality. For channel estimation, if real-time feedback of the channel quality is required, the value of $window\_size$ should be appropriately small, and if more accurate channel quality estimation is required, the value of $window\_size$ should be appropriately large.

We use the array $total\_packet\_oks[37]$ to record the total number of successfully transmitted packets for each channel. When a packet is successfully transmitted on channel $i$, $total\_packet\_oks[i]$ is incremented by one. Similarly, we use $total\_packet\_errors[37]$ to record the number of packet transmission failures for each channel, incrementing $total\_packet\_errors[i]$ when a failure occurs.

The average PDR of channel $i$ over the entire communication phase is computed using Equation~\eqref{eq:pdr_total}:

\begin{equation}
\resizebox{0.44\textwidth}{!}{%
$ PDR\_total[i] = \frac{total\_packet\_oks[i]}{total\_packet\_errors[i] + total\_packet\_oks[i]} \times 100\% $
}
\label{eq:pdr_total}
\end{equation}

\subsubsection{Parameter Analysis}
In BLE, the factors that affect channel quality and packet delivery rate mainly include environmental interference such as disturbances from other wireless devices, the distance between devices since greater distances cause more severe signal attenuation, transmission power and reception sensitivity which influence the communication distance and quality between devices, the PHY mode of transmission, and channel congestion, where excessive device communication in certain frequency bands can lead to packet loss or delays. These factors work together to determine the communication efficiency and stability of BLE devices. This section discusses the impact of transmission power, received signal strength indication, physical layer mode, and transmission channel on channel quality.  
\paragraph{Transmission Power (TXP)}

\begin{figure}[h]
    \centering
    \includegraphics[width=0.5\textwidth]{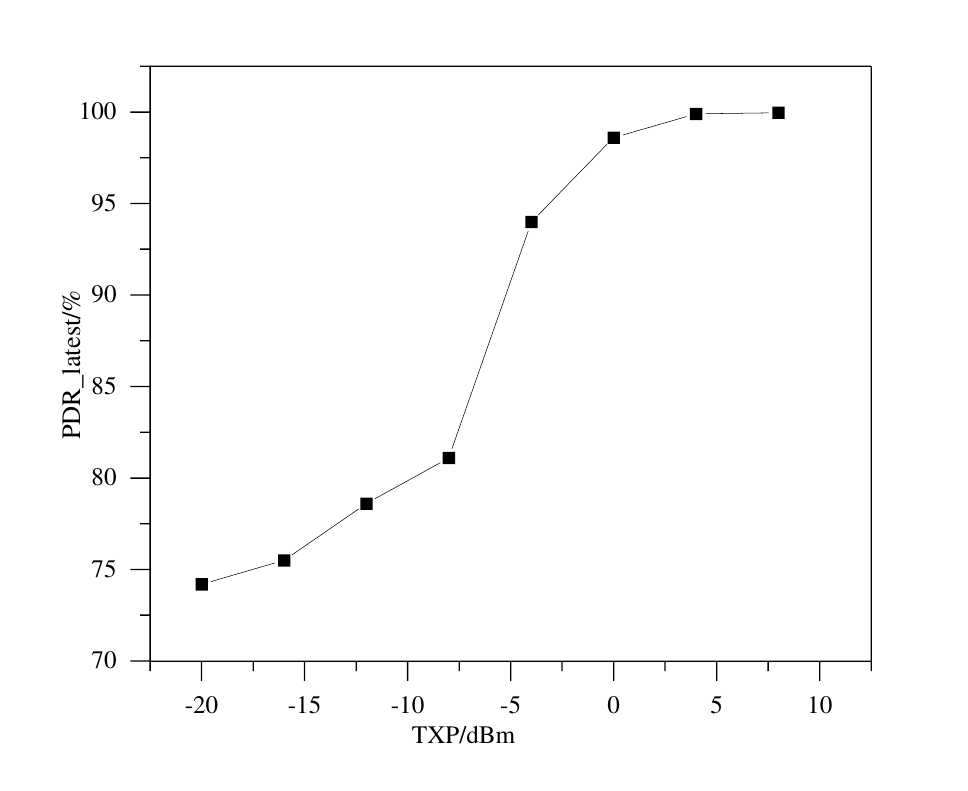}
    \caption{Effect of Transmit Power $TXP$ on $PDR$ of $BLE$ Channel Quality}
    \label{fig:flowchart5}
\end{figure}

Bluetooth transmission power refers to the power level used by a Bluetooth device when transmitting a wireless signal. It determines the strength of the signal transmitted by the Bluetooth device, thus affecting the transmission range and reliability of the signal. Transmission power is usually expressed in decibels milliwatts (dBm). Figure \ref{fig:flowchart5} shows how $PDR\_latest$ changes with an increase in TXP. 

In the initial stage, as the transmission power increases, the PDR may increase relatively quickly, because the increased signal strength can overcome some of the interference in transmission, thereby increasing the likelihood of successful packet transmission. As the transmission power continues to increase, the growth rate of the PDR will gradually slow down, and it may even reach a saturation state. In some cases, increasing the transmission power will no longer significantly increase the PDR, and may result in additional power consumption and interference.

\begin{figure}[h]
    \centering
    \includegraphics[width=0.5\textwidth]{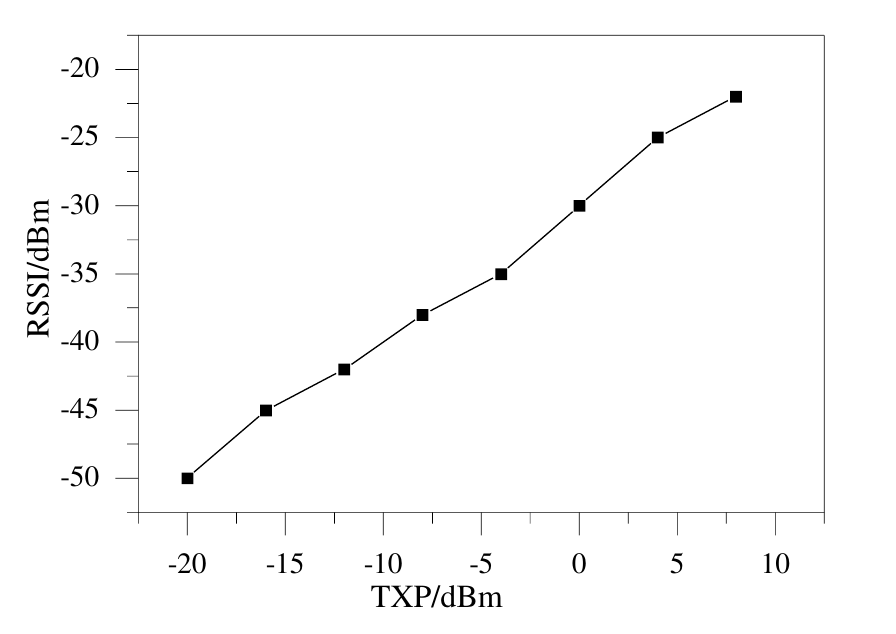}
    \caption{Relationship between Transmit Power TXP and Received Signal Strength RSSI}
    \label{fig:flowchart6}
\end{figure}

\paragraph{Received Signal Strength Indicator (RSSI)} 
Received Signal Strength Indication (RSSI)\cite{ref12} is an indicator that measures the strength of the received signal and is commonly used to evaluate the signal quality in wireless communication systems. It indicates the power level of the received signal in dBm (decibel milliwatts). In general, the stronger the received signal, the higher the RSSI value, and vice versa. Figure \ref{fig:flowchart6} shows the relationship between RSSI and TXP under PHY 2M. As can be seen from the figure, in a wireless communication system, the RSSI value is usually positively correlated with the transmission power. This means that the greater the transmission power, the greater the received signal strength will usually be, resulting in a higher RSSI value. To a certain extent, RSSI is linearly correlated with TXP.

\begin{figure}[h]
    \centering
    \includegraphics[width=0.5\textwidth]{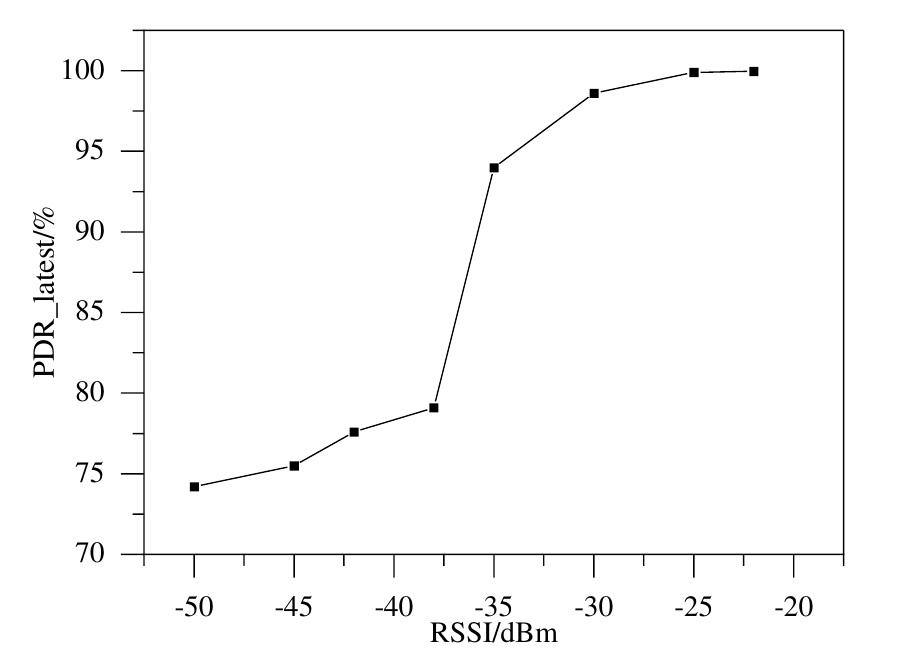}
    \caption{Effect of RSSI on PDR in different PHY modes}
    \label{fig:flowchart7}
\end{figure}

Figure \ref{fig:flowchart7} shows the relationship between RSSI and $PDR\_latest$ under PHY 2M. RSSI shows an increasing trend with an increase in PDR. However, this relationship is not linear, but shows a nonlinear growth trend, and the changing trend is almost the same as the change between PDR and TXP in Figure \ref{fig:flowchart5}. Transmit power is not convenient for real-time monitoring in FM systems, while RSSI is convenient for real-time monitoring, and RSSI is linearly related to TXP to a certain extent. Therefore, when considering transmission power adaptation, RSSI can be used as a reference index value.
\paragraph{Physical Layer Mode (PHY)} 
Before Bluetooth 5.0 , BLE only supported PHY 1M\cite{ref13}with a transmission rate of 1 Mbps. However, Bluetooth 5.0 introduces high-speed (PHY 2M) and long-range (PHY Coded) wireless transmission modes. Using PHY 2M enables higher transmission rates (2 Mbps), which means that data can be transmitted faster and the device can enter sleep mode more quickly to save more power, or the extra time can be used to send more data, effectively doubling the throughput of the BLE connection. However, the price to pay for this is a slightly shorter range. Using PHY Coded, on the other hand, can significantly increase the transmission range, but at the cost of reduced throughput\cite{ref14}.

\begin{figure}[h]
    \centering
    \includegraphics[width=0.5\textwidth]{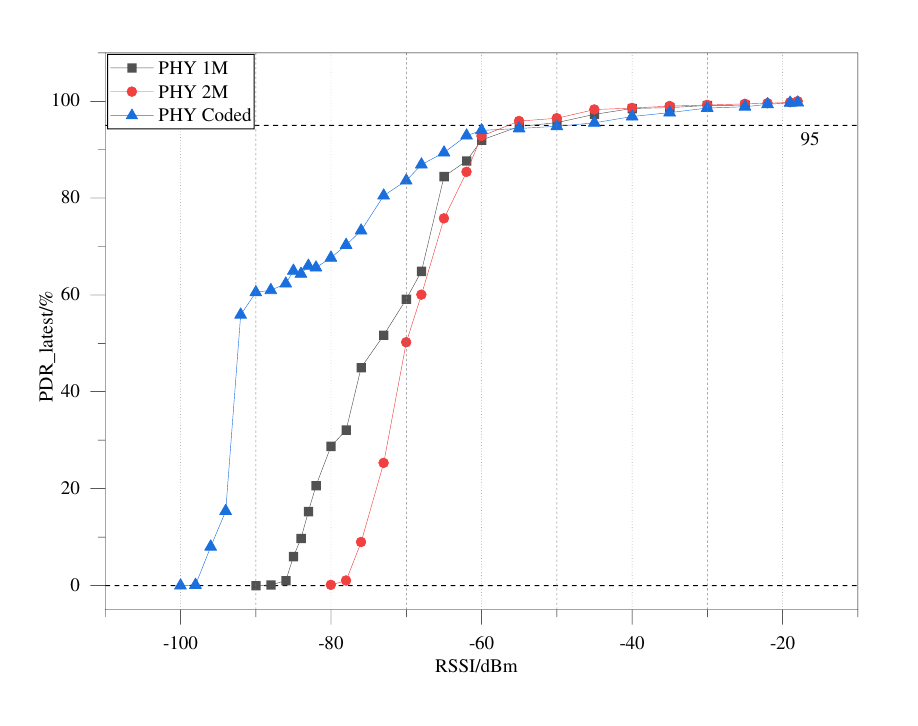}
    \caption{Effect of PHY modes on PDR}
    \label{fig:flowchart8}
\end{figure}

Figure \ref{fig:flowchart8} shows the packet delivery ratio (PDR\_latest) of PHY 1M, PHY 2M and PHY Coded with varying RSSI when the BLE peripheral device is gradually moved away from the central device under the transmission power TXP = 4 dBm and indoor corridor environment. As can be seen from Figure \ref{fig:flowchart8}, the lower limit of RSSI for PHY 2M is about -78 dBm, for PHY 1M is about -90 dBm, and for PHY Coded is about -100 dBm. When RSSI is less than -55 dBm, the packet delivery rate of the three PHY modes is almost the same. However, PHY 2M has advantages in transmission rate and power consumption. PHY 1M has certain advantages in speed and power consumption over PHY Coded when it is between -55 dBm and -65 dBm. At less than -65 dBm to -70 dBm, PHY Coded has better communication quality, although it consumes more power than PHY 1M and PHY 2M. Therefore, when performing PHY adaptive adjustment, the RSSI and TXP can be adjusted together as parameters. This adjustment can help balance the relationship between communication range, power consumption, and interference, thereby ensuring good performance and stability of the communication system while minimizing power.

In Figure \ref{fig:flowchart8}, the intersection point appears at an RSSI of about -55 dBm. When the RSSI is greater than -55 dBm, PHY Coded, PHY 1M, and PHY 2M all achieve a relatively high packet delivery rate ($PDR\_latest$ \textgreater 95{\%}). When the RSSI is high, the PDR of different PHY modes will be similar, because a high RSSI generally indicates a high received signal strength, which is conducive to reducing the probability of transmission errors. Therefore, to a certain extent, the transmission error rates of PHY Coded, PHY 1M and PHY 2M will tend to be similar when the RSSI is high.

\paragraph{Transmission Channel} 
BLE uses 40 channels in the 2.4 GHz ISM band for communication, each with a channel spacing of 2 MHz. Three of these channels are broadcast channels, on which BLE devices regularly send advertising packets for device discovery and connection. Normally, a BLE device broadcasts on one of the three broadcast channels at random. These broadcast channels have fixed frequencies, 2402 MHz, 2426 MHz and 2480 MHz, respectively. The remaining 37 channels (channel $0$ to channel $36$) are used for data transmission. Data channels are used for actual data transmission. The frequency range of these data channels spans from  2404 MHz to 2480 MHz., with an interval of 2 MHz. The 37 channels of BLE are managed using a channel map\cite{ref15}, which is a mechanism for determining which channel a data packet is sent on in BLE communication. The channel map is represented using a 37-bit bitmap, where each bit corresponds to a BLE channel. If the bit is 1, the channel is available; if the bit is 0, the channel is disabled. The array $channel_map=[0xff, 0xff, 0xff, 0xff, 0x1f]$ is used to represent the BLE channel map, where each byte in this array is used to represent 8 BLE channels. For example, “$0xff$” represents a byte in which all 8 bits are set to 1. In binary notation, this byte is “$11111111$”, which means that all 8 channels corresponding to this byte are available.

\begin{figure}[h]
    \centering
    \includegraphics[width=0.48\textwidth,trim=0 0 0 0, clip]{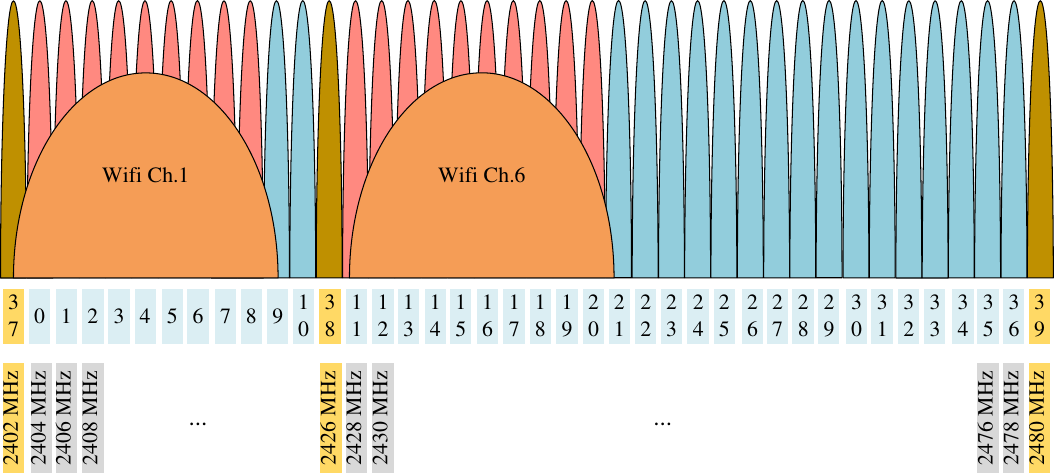}
    \caption{Bluetooth channels under co-channel Wi-Fi interference}
    \label{fig:flowchart5.7}
\end{figure}

In practice, Bluetooth is more susceptible to interference from Wi-Fi signals in the 2.4 GHz band. As shown in Figure \ref{fig:flowchart5.7}, when the 2.4 GHz Wi-Fi channel is set to channel 1, the affected Bluetooth channels (ch) are ch0 to ch8, and when the 2.4 GHz Wi-Fi channel is set to ch6, the affected Bluetooth channels are ch11 to ch20.

\begin{figure}[h]
    \centering
    \includegraphics[width=0.5\textwidth]{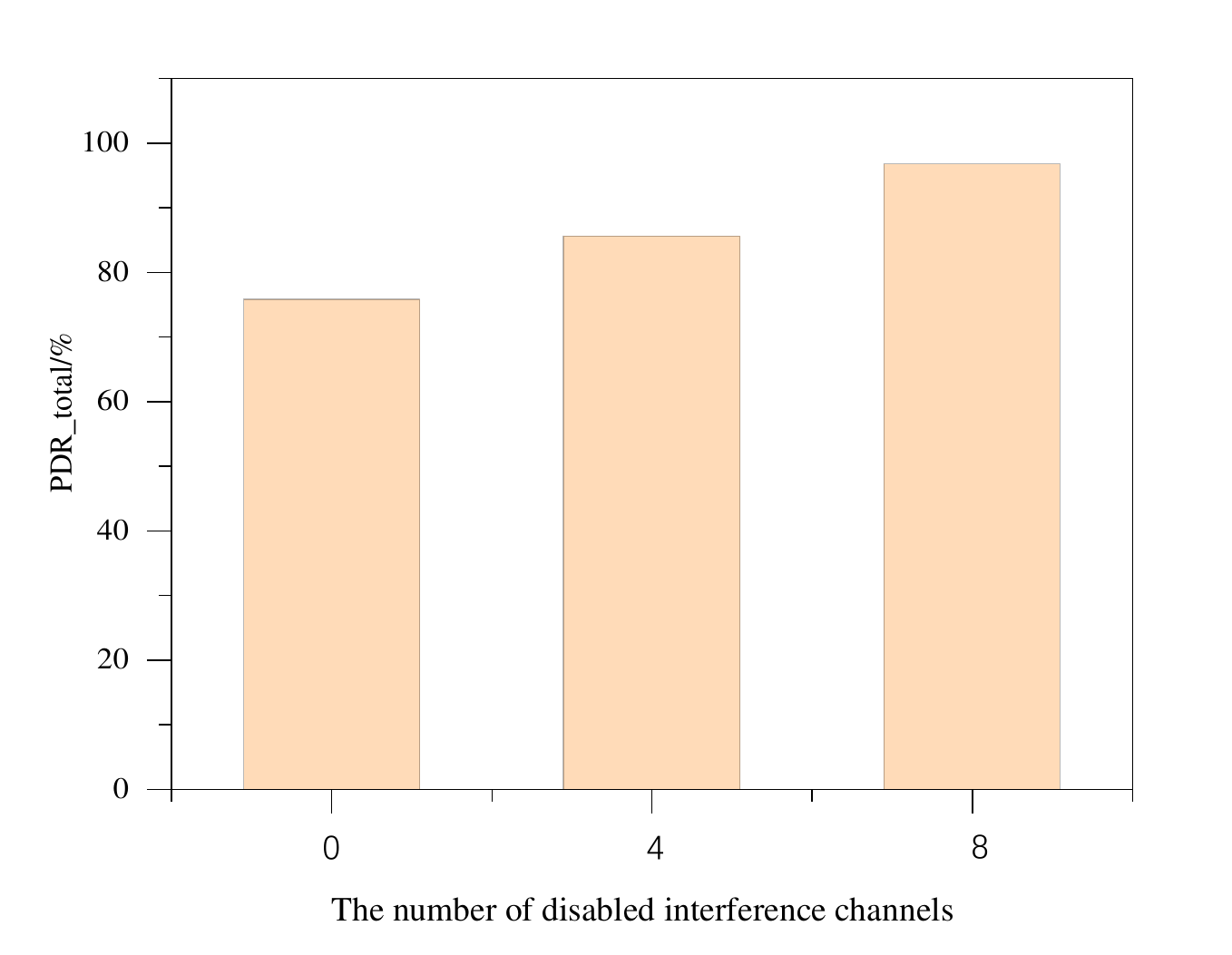}
    \caption{Effect of disabling interference channels on channel quality}
    \label{fig:flowchart5.8}
\end{figure}

In order to explore the impact of Bluetooth low energy channel interference on the communication quality PDR, a 2.4 GHz Wi-Fi router was used to interfere with the Bluetooth transmission. The Wi-Fi channel was set to channel 1, the signal strength was -25 dBm, the average download speed of the router was 8 MB/s, and the Bluetooth PHY mode is PHY 2M. Figure \ref{fig:flowchart5.8} shows the average value of $PDR\_total$ in one minute with all channels enabled, four interfering channels disabled, and eight interfering channels disabled under Wi-Fi interference. This shows that by real-time and reasonable monitoring of the quality of the communication channel, interfering channels can be disabled and removed in time to significantly improve the quality of Bluetooth transmission.

\subsection{Scheme Design}
Based on the analysis results of the BLE channel quality parameters in Section 5.1, this section proposes an adaptive secure communication scheme and designs adaptive frequency-hopping, adaptive PHY mode and transmission power algorithms to improve the overall channel quality and channel resource utilization of Bluetooth low energy.
\subsubsection{Adaptive Frequency-Hopping Algorithm}

\begin{algorithm}[htbp]
    \caption{Adaptive Frequency-Hopping Algorithm}
    \label{alg:adaptive_freq}
    \renewcommand{\algorithmicrequire}{\textbf{Input:}}
    \renewcommand{\algorithmicensure}{\textbf{Output:}}
    \begin{algorithmic}[1]
        \REQUIRE 
        
        $PDR\_latest$,
        
        $PDR\_threshold$, 
        
        $last\_channel\_map[5]$, 
        
        $new\_channel\_map[5]$, 
        
        $used\_channel$, 
        
        $channel\_threshold$
        \ENSURE 
        
        $new\_channel\_map[5]$
        
        \STATE Initialize the variable $channel\_map[5]$
        \STATE Calculate the number of channels in $used\_channel$ and $last\_channel\_map[5]$
        \IF{$used\_channel < channel\_threshold$}
            \STATE Update the channel map as follows: $init\_channel\_map[5] = \{0xFF, 0xFF, 0xFF, 0xFF, 0xFF\}$ (all channels are available)
        \ENDIF
        \FOR{$j = 0$ to 4}
            \STATE Set $init\_channel\_map[j] = last\_channel\_map[j]$
        \ENDFOR
        
        \FOR{$j = 0$ to 36}
            \IF{$PDR\_latest[j] < PDR\_threshold$}
                \STATE Update the new channel map: $new\_channel\_map[5] = init\_channel\_map[5]$
            \ENDIF
        \ENDFOR
        
        \RETURN $new\_channel\_map$
    \end{algorithmic}
    \label{A1}
\end{algorithm}

This section proposes an adaptive frequency-hopping algorithm. The core idea is to evaluate the channel quality in real time. By continuously monitoring the PDR of each channel, channels with a PDR below a preset threshold are automatically identified and blacklisted. This avoids interference and improves the reliability of data transmission. In addition, a channel restart mechanism is designed to allow the system to re-evaluate and potentially restore the use of previously blacklisted channels at the appropriate time.

The Algorithm \ref{A1} checks whether the number of channels that were active during the last event is less than a preset threshold. If it is less than the threshold, all channels will be re-enabled at this time to expand the number of channels and increase the system's range of channel selection, which is conducive to finding better channels. On the premise of ensuring that the number of channels meets the requirements, the algorithm will continue to select better channels based on the channels used in the last event. By removing channels with $PDR\_latest$ less than a preset threshold, channels with poor quality are dynamically eliminated, thereby improving the quality of the system's channel selection and improving the stability and reliability of communication.
\subsubsection{Adaptive TXP and PHY Mode Control algorithm}

This section proposes a PHY and power adaptive selection algorithm that adapts the PHY mode configuration and transmission power based on real-time readings of the RSSI. By comprehensively considering the current channel conditions and RSSI data, the most suitable PHY mode and appropriate transmission power level can be dynamically selected to improve the overall channel quality of BLE and optimize communication performance and energy efficiency.

The Algorithm \ref{A2} improves communication quality by dynamically adjusting the PHY mode and transmission power. Based on real-time RSSI values and the communication environment, the algorithm selects the most suitable PHY mode and transmission power to ensure the stability and reliability of communication. The transmission power is increased or decreased according to actual needs to balance the relationship between communication range and power consumption. The algorithm achieves the widest range of communication through the initial setting of PHY Coded and a higher transmission power, and then gradually adjusts according to the real-time RSSI value. This dynamic adjustment can adapt to different communication environments and needs to ensure that the system maintains a stable communication connection in various situations.

\begin{algorithm}[!h]
    \caption{Adaptive TXP and PHY Mode Control Algorithm}
    \label{alg:adaptive_power_phy}
    \renewcommand{\algorithmicrequire}{\textbf{Input:}}
    \renewcommand{\algorithmicensure}{\textbf{Output:}}
    \begin{algorithmic}[1]
        \REQUIRE $RSSI\_current$, $PDR\_current$, 
        
        $TXP\_current$, $PHY\_current$; 
        
        $RSSI\_high$, $RSSI\_low$; 
        
        $PDR\_high$, $PDR\_low$; 
        
        \ENSURE $PHY\_new$, $TXP\_new$; 
        
        \STATE Initialize $PHY$, $TXP$: $PHY$ = PHY Coded, $TXP$ = 8 dBm 
        
        \IF{$RSSI\_current$ $>$ $RSSI\_high$ \AND $PDR\_current$ $>$ $PDR\_high$}
            \IF{$TXP\_current$ $>$ -20}
                \STATE $TXP\_new$ = $TXP\_current$ - 4 
            \ELSE
                \IF{$PHY\_current$ $\neq$ PHY 2M}
                    \STATE $PHY\_new$ = PHY 2M 
                \ENDIF
            \ENDIF
        \ELSIF{$RSSI\_current$ $<$ $RSSI\_low$ \AND $PDR\_current$ $<$ $PDR\_low$}
            \IF{$TXP\_current$ $<$ 8}
                \STATE $TXP\_new$ = $TXP\_current$ + 4 
            \ELSE
                \IF{$PHY\_current$ $\neq$ PHY Coded}
                    \STATE $PHY\_new$ = PHY Coded 
                \ENDIF
            \ENDIF
        \ELSE
            \STATE $PHY\_new$ = $PHY\_current$, $TXP\_new$ =  $TXP\_current$ 
        \ENDIF
        
        \RETURN $PHY\_new$, $TXP\_new$
    \end{algorithmic}
    \label{A2}
\end{algorithm}

\section{Lightweight Vehicle-Key Authentication Protocol}\label{666}
\subsection{Scheme Design}

In the current RKE system, the inherent characteristics of unidirectional radio frequency communication limit the ability to directly authenticate the identity of the car key and the signal, which exposes the vehicle to security threats such as replay attacks and message tampering. To overcome these security threats and improve the security of the system, this paper proposes a lightweight car key authentication protocol. The protocol is based on the efficient Elliptic Curve Digital Signature Algorithm (ECDSA) to achieve strong authentication and uses Advanced Encryption Standard (AES) symmetric encryption technology to ensure the confidentiality and integrity of the signal.

\begin{figure*}[h]
    \centering
    \includegraphics[width=1\textwidth]{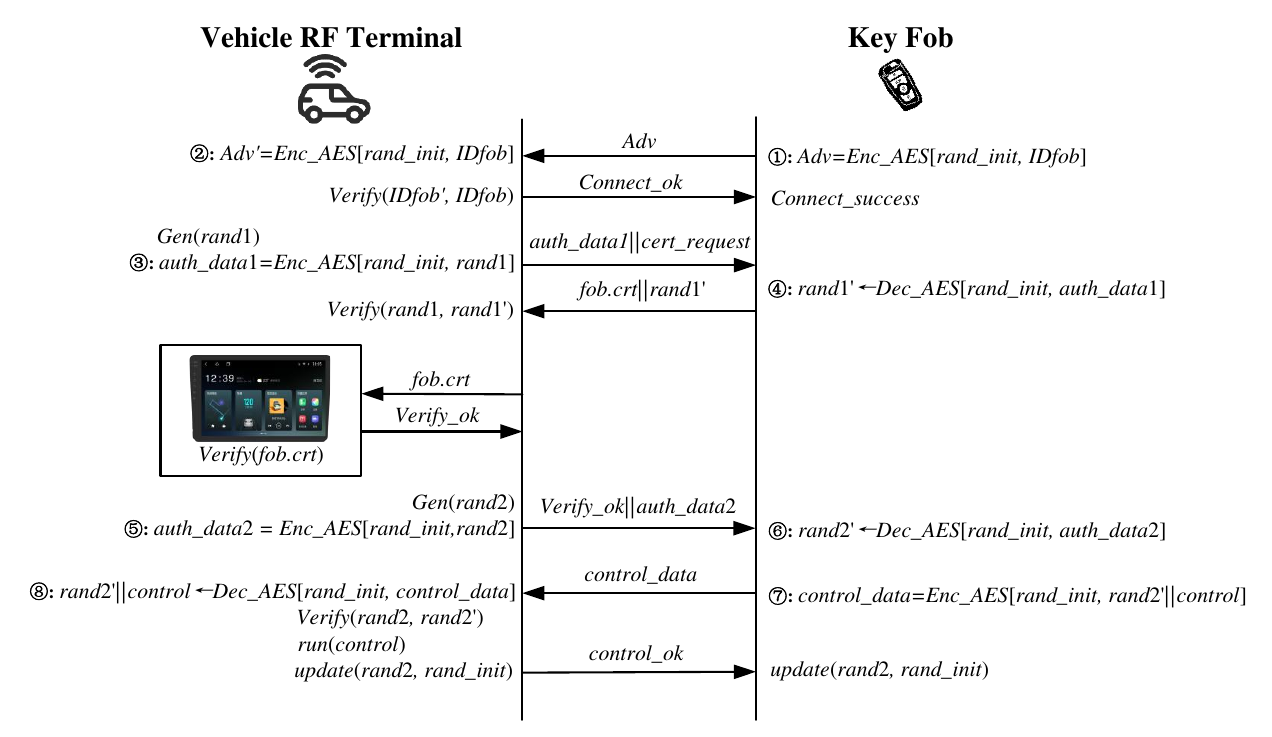}
    \caption{Vehicle-Key Authentication Protocol}
    \label{111}
\end{figure*}

The authentication process has three stages, as shown in the Figure \ref{111}: the communication establishment stage, the authentication stage, and the encrypted transmission control command stage. Breaking it down further, it can be divided into ten steps.

\textbf{\subsubsection{Communication Establishment Stage}}
\textbf{Step 1}: The key fob encrypts the $PIDfob$ using the AES algorithm as described in Equation \ding{172} of Figure \ref{111}, with the encryption key being $rand\_init$. The resulting cipher text, denoted as $Adv$, is broadcasted by the key fob.

\textbf{Step 2}: The key fob scans the broadcast, and if the address matches one in the whitelist, it encrypts $PIDfob'$ using the AES algorithm as described in Equation \ding{173} of Figure \ref{111}, with the encryption key being $rand\_init$. The resulting cipher text, denoted as $Adv'$, is generated. The vehicle RF terminal verifies $Adv'$ and checks if it matches $Adv$. If they are equal, a communication connection is established.

\textbf{\subsubsection{Authentication Stage}}
\textbf{Step 3}: The vehicle RF terminal generates a random number $rand1$ to prevent message replay. It then encrypts $rand1$ using the AES algorithm, as described in Equation \ding{174} of Figure \ref{111}, with the encryption key being $rand\_init$. The resulting authentication data packet, denoted as $auth\_data1$, is created. The vehicle RF terminal sends the authentication data packet $auth\_data1$ and the certificate request $cert\_request$ to the key fob, initiating the authentication request.

\textbf{Step 4}: The key fob receives the certificate request $cert\_request$ and decrypts the authentication data packet $auth\_data1$ using the AES algorithm, as described in Equation \ding{175} of Figure \ref{111}, with the decryption key being $rand\_init$, resulting in $rand1'$. The public key certificate, which is publicly available, does not require encryption before being sent. The key fob then sends the key certificate $fob.crt$ and the random number $rand1'$.

\textbf{Step 5}: The vehicle RF terminal receives $fob.crt$ and $rand1'$.  

    a. It checks whether $rand1' = rand1$. If not, it sends an authentication failure signal:  $"verify\_failed"$ with the reason $"error\ rand1"$ .
        
    b. If $rand1' = rand1$, it forwards $fob.crt$ to the vehicle system for step 6.

\textbf{Step 6}: The vehicle certificate verification system uses ECDSA signature verification to authenticate the key fob certificate and checks whether the certificate’s $PIDfob$ matches the decrypted $PIDfob$. 

   a. The vehicle certificate verification system successfully authenticates the certification of key fob and notifies the vehicle RF terminal of the authentication success. The vehicle RF terminal generates a new random number $rand2$ to prevent replay attacks. It then encrypts the random number $rand2$ using the AES algorithm with the initial symmetric key $rand\_init$, as described in Equation \ding{176} of Figure \ref{111}, resulting in the encrypted authentication data packet $auth\_data2$. The vehicle RF terminal sends the authentication data packet $auth\_data2$ and the certificate authentication success flag $"verify\_ok"$.

   b. If verification fails, it sends $"verify\_failed"$ with reason $"crt\ error"$.  

\textbf{Step 7}: The key fob processes the received data.  

   a. If `"verify\_failed"` is received, it enters sleep mode until the next wake-up.  
   
   b. If `"verify\_ok"` is received, authentication is successful, and step 8 is executed.  

\textbf{\subsubsection{Encrypted Transmission of Control Commands Stage}}
\textbf{Step 8}: Upon successful authentication, the key fob decrypts using the AES algorithm as described in Equation \ding{177} of Figure \ref{111} to obtain $rand2'$. Then, using Equation \ding{178} of Figure \ref{111}, the key fob encrypts the control command $control$ and the random number $rand2'$ with the initial symmetric key $rand\_init$, resulting in the encrypted control instruction packet $control\_data$. The key fob then sends the encrypted control instruction packet $control\_data$ to the vehicle RF terminal.

\textbf{Step 9}: The vehicle RF terminal receives the encrypted control instruction packet $control\_data$ and decrypts it using the AES algorithm as described in Equation \ding{179} of Figure \ref{111} to obtain the random number $rand2'$ and the control command $control$. It then checks whether $rand2'$ is equal to the previously generated $rand2$.

   a. If $rand2' = rand2$, it executes the command (e.g., unlocking the door) and sends $"control\_ok"$. It also updates $rand\_init$ using $rand2$ as the new session key to ensure unique encryption keys for each authentication process.  
   
   b. If $rand2' \neq rand2$, the command is not executed, and the connection is terminated.  

\textbf{Step 10}: The key fob receives $"control\_ok"$, confirming the successful execution of the command. It then updates $rand\_init$ using $rand2$ as the new session key for future communications.  

Through this protocol, the vehicle and key fob establish one-way authentication, ensuring that only an authorized key fob can communicate with the vehicle. Once authentication is complete, AES encryption is used to secure control command transmissions. Furthermore, updating the session key in each authentication cycle prevents unauthorized access and key leakage.  

\section{Performance Evaluation} 
In order to verify the effectiveness and security of the proposed adaptive frequency-hopping RKE secure communication scheme, this paper implements an RKE communication system based on adaptive frequency-hopping on two Bluetooth nRF 52840\cite{ref50} embedded development boards and two PCs based on an embedded software and hardware development environment. The proposed scheme is experimentally verified and tested and evaluated from three aspects: functional testing, performance testing, and security testing.

\subsection{Experiment Setup}
The development environment is configured to support efficient implementation and testing of the prototype system. As shown in Table \ref{tab:dev_environment}, we utilize both Windows and Ubuntu for flexibility. OpenSSL is used for secure certificate authentication, while Visual Studio Code and the nRF Connect SDK provide a robust platform for embedded development in C. This setup enables seamless integration and testing of the authentication system on the target hardware.

\begin{table}[h]
    \centering
    \caption{Development Environment for the Prototype System}
    \label{tab:dev_environment}
    \begin{tabular}{|c|c|}
        \hline
        \textbf{Component} & \textbf{Details} \\
        \hline
        Processor & Intel Core i7-10700 (2.90 GHz) \\
        RAM & 16.0 GB \\
        Operating Systems & Windows 10 (x64), Ubuntu 20.04 (via VM) \\
        Certificate Authentication & OpenSSL \\
        Development Tool & Visual Studio Code \\
        Programming Language & C \\
        SDK & nRF Connect SDK v2.5.0 \\
        Toolchain & nRF Connect Toolchain v2.5.0 \\
        \hline
    \end{tabular}
\end{table}

\begin{figure}[h]
    \centering
    \includegraphics[width=0.5\textwidth]{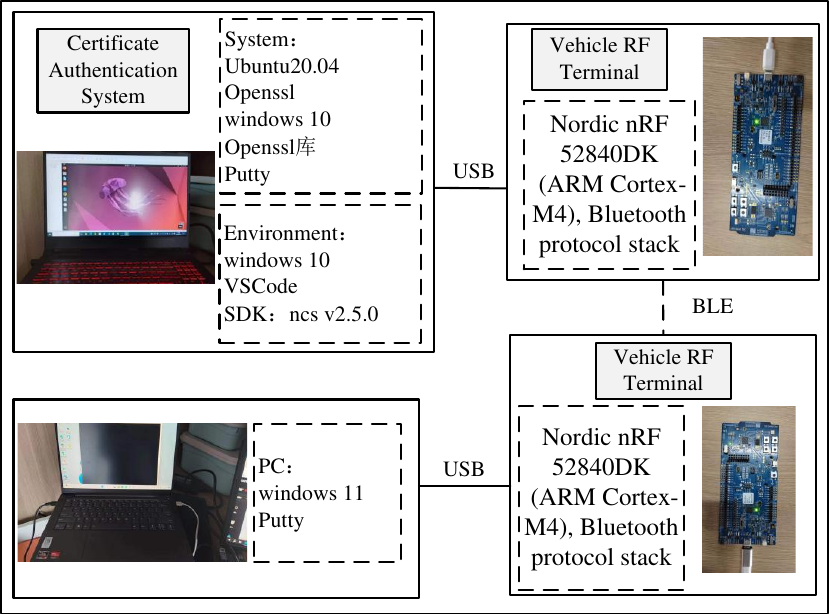}
    \caption{Test environment components}
    \label{fig:flowchart6.1}
\end{figure}

The composition and configuration of the test environment are shown in Figure \ref{fig:flowchart6.1}. It mainly consists of a key end, a vehicle RF end, and an authentication system. The key end consists of an nRF 52840 and a PC, which displays the output information of the running program. The vehicle RF end consists of an nRF 52840 and another PC, which debugs the embedded system and displays the output information of the running program. It also serves as the vehicle certificate authentication system to authenticate the certificate.

\subsection{Adaptive Communication Scheme Analysis}
\textbf{\subsubsection{The impact of static Wi-Fi interference on channel quality}}

In order to explore the improvement of channel quality brought about by the proposed adaptive secure communication system, the changes of real-time channel quality $PDR\_latest$ over time under two conditions were selected: a mild indoor Wi-Fi interference environment and a stronger indoor Wi-Fi interference environment, by controlling the transmission power and distance of the interference source. The changes of real-time channel quality $PDR\_latest$ over time for PHY 2M, PHY 1M, and PHY Coded without adaptive frequency hopping were also analyzed. The adaptive frequency hopping communication scheme selects a $window\_size$ length of 25 and a PDR filtering threshold of 95\%.

\paragraph{Mild WiFi Interference}

The experimental environment in this section is an indoor  Wi-Fi environment with light interference, in which the strongest interfering signal source is about -55 dBm, and all other signal sources are  below -60 dBm. The Wi-Fi download speed remains around 10 M/s. 

\begin{figure}[h]
    \centering
    \includegraphics[width=0.48\textwidth]{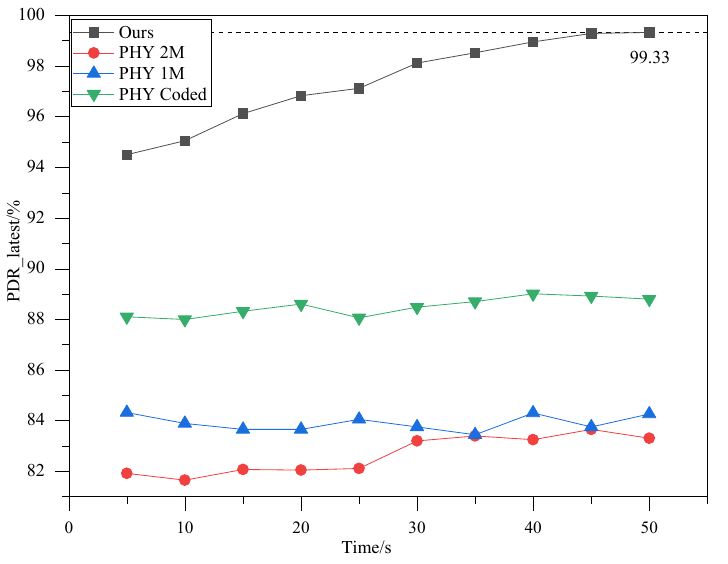}
    \caption{Time-dependent change of $PDR\_latest$ under mild interference conditions}
    \label{fig:light_interference_pdr}
\end{figure}

Figure~\ref{fig:light_interference_pdr} illustrates the variation of $PDR\_latest$ over time under light interference conditions for the adaptive communication scheme and the non-adaptive scheme using PHY 2M, PHY 1M, and PHY Coded.

\paragraph{Strong WiFi Interference}

The experimental environment in this section is an indoor  Wi-Fi environment with strong interference, in which the strongest interference signal source is about -25 dBm, and all other signal sources are below -50 dBm. The Wi-Fi download speed remains around 10 M/s.

\begin{figure}[h]
    \centering
    \includegraphics[width=0.48\textwidth]{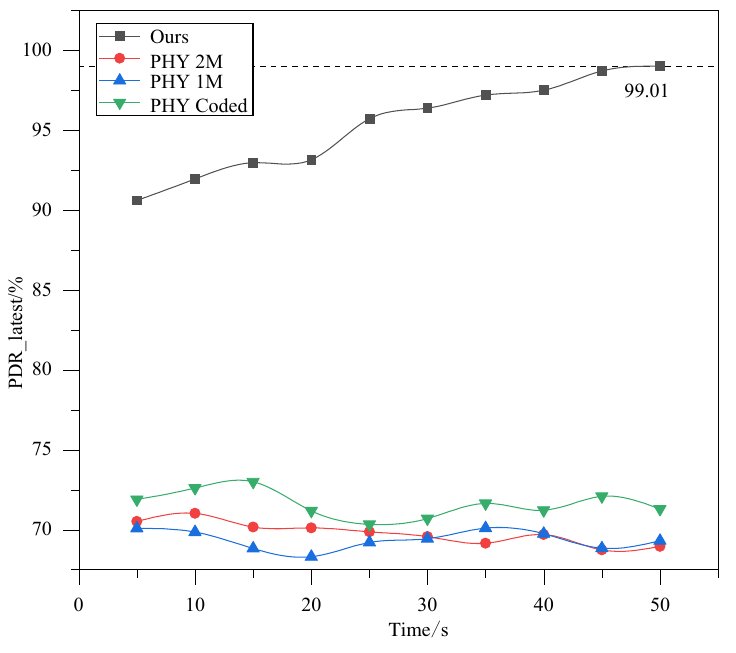}
    \caption{Time-dependent change of $PDR\_latest$ under strong interference conditions}
    \label{fig:strong_interference_pdr}
\end{figure}

Figure~\ref{fig:strong_interference_pdr} depicts the variation of $PDR\_latest$ over time under strong interference conditions for the adaptive frequency hopping communication scheme, PHY 2M, PHY 1M, and PHY Coded.

From Figures~\ref{fig:light_interference_pdr} and \ref{fig:strong_interference_pdr}, it can be seen that the proposed adaptive frequency-hopping scheme effectively improves Bluetooth communication quality. 

Under light interference conditions, the PDR of PHY Coded remains around 80\%, while the real-time channel quality $PDR\_latest$ of PHY 1M and PHY 2M remains between 70\% and 73\%. In contrast, the PDR of the adaptive scheme continuously increases from 93\% to 99.23\% and remains stable.

Under strong interference conditions, the PDR of PHY Coded, PHY 1M, and PHY 2M remains between 70\% and 73\%. The PDR of the adaptive scheme increases from approximately 85\% to 99.01\% and remains stable.

\subsubsection{The impact of dynamic Wi-Fi interference on channel quality}

In order to test the performance of the proposed adaptive secure communication system in a dynamic environment, the overall channel quality $PDR\_{total}$ was tested over time by switching Wi-Fi channels to compare the performance with and without adaptive communication.

In the dynamic Wi-Fi interference test, the signal strength of the Wi-Fi with the strongest interference is approximately -55 dBm, the download speed is about 10 MB/s, and the distance is around 50 cm. Initially, the Wi-Fi channel is set to ch1 (channel 1), and the Wi-Fi channel is switched to ch5, ch10, and ch13 at 50 s, 100 s, and 150 s, respectively. 

\begin{figure}[h]
    \centering
    \includegraphics[width=0.48\textwidth]{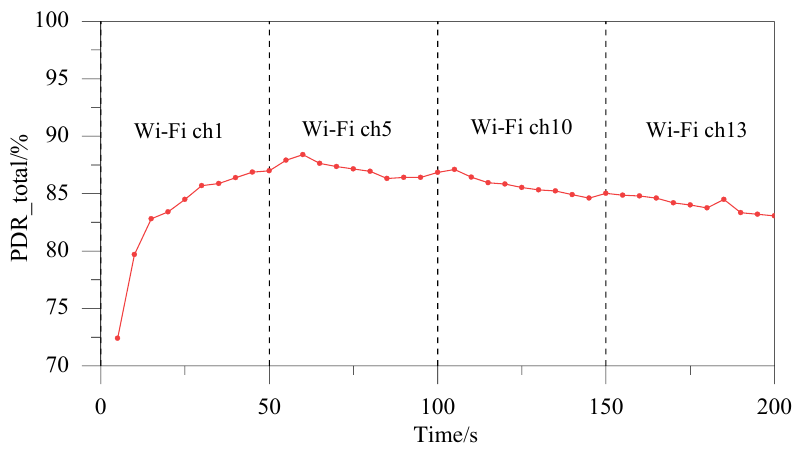}
    \caption{Dynamic change of $PDR\_total$ over time under Wi-Fi interference without adaptive communication system}
    \label{fig:non_adaptive_dynamic}
\end{figure}

\begin{figure}[h]
    \centering
    \includegraphics[width=0.48\textwidth]{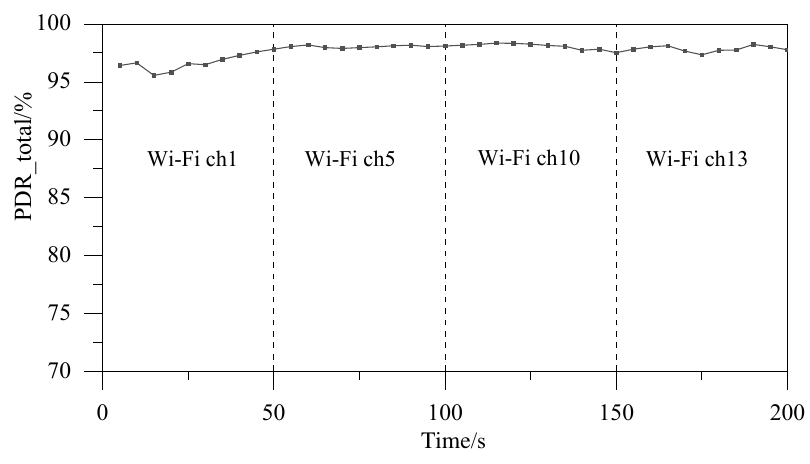}
    \caption{Dynamic change of $PDR\_total$ over time under Wi-Fi interference with adaptive communication system}
    \label{fig:adaptive_dynamic}
\end{figure}

Figure~\ref{fig:non_adaptive_dynamic} shows the variation of $PDR\_{total}$ over time for the non-adaptive communication system under dynamic Wi-Fi interference, while Figure~\ref{fig:adaptive_dynamic} presents the variation of $PDR\_{total}$ over time for the adaptive communication system under the same conditions.

From Figures~\ref{fig:non_adaptive_dynamic} and \ref{fig:adaptive_dynamic}, it can be observed that when the Wi-Fi channel is dynamically switched, the $PDR\_{total}$ of the non-adaptive communication system increases from 73\% to approximately 87\%, then gradually decreases and stabilizes between 82\% and 85\%. In contrast, the $PDR\_{total}$ of the adaptive communication system remains stable between 95\% and 98\%. With dynamic Wi-Fi channel switching, the overall communication quality of the adaptive communication system is significantly improved compared to the non-adaptive communication system.

From the above experiments, it can be concluded that the adaptive communication module can significantly enhance the PDR and improve the quality of the wireless communication channel. This is achieved by adaptively selecting an optimal channel, PHY mode, and transmission power under varying Wi-Fi interference strengths and dynamic Wi-Fi interference conditions.

\subsubsection{Memory Overhead Analysis}

In order to verify the usability and additional memory cost of the proposed adaptive frequency hopping key fob secure communication scheme, the proposed adaptive algorithm was implemented on the BLE Bluetooth embedded chip nRF 52840. 

\begin{table}[h]
    \centering
    \caption{Memory Overhead Without Adaptive Communication System}
    \begin{tabular}{|c|c|c|c|}
        \hline
        Memory Region & Used Size & Total Size & Usage Percentage \\
        \hline
        FLASH & 236,212 B & 1 MB & 22.54\% \\
        RAM & 39,916 B & 256 KB & 15.23\% \\
        IDT\_LIST & 0 GB & 2 KB & 0.00\% \\
        \hline
    \end{tabular}
    \label{table:non_adaptive}
\end{table}

\begin{table}[h]
    \centering
    \caption{Memory Overhead With Adaptive Communication System}
    \begin{tabular}{|c|c|c|c|}
        \hline
        Memory Region & Used Size & Total Size & Usage Percentage \\
        \hline
        FLASH & 289,152 B & 1 MB & 27.58\% \\
        RAM & 44,376 B & 256 KB & 16.93\% \\
        IDT\_LIST & 0 GB & 2 KB & 0.00\% \\
        \hline
    \end{tabular}
    \label{table:adaptive}
\end{table}

As can be seen from Tables~\ref{table:non_adaptive} and \ref{table:adaptive}, the RAM overhead of the non-adaptive communication system is 39,916 B and the FLASH overhead is 236,212 B; the RAM overhead of the adaptive frequency hopping system is 44,376 B and the FLASH storage overhead is 289,152 B. Compared with the non-adaptive frequency hopping system, the adaptive frequency hopping system has an additional 14\% memory control overhead and an additional 18\% storage overhead.

\subsubsection{Power Overhead Analysis}

In order to test the additional power consumption required by the proposed adaptive communication system to monitor channel quality in real time, the adaptive PHY mode and TXP adaptation modules were turned off (different PHY modes have different transmission powers). In PHY 2M mode, the transmission power is 0 dBm, and the RSSI is about -40 dBm. The power consumption and average power consumption of the system with and without adaptive communication systems were tested using IoT power within 10 s of operation. The additional power consumption at this time is the power consumption of the real-time monitoring channel.

Through experiments,the power consumption of the non-adaptive frequency hopping system BLE is 57.27~$\mu$Ah and the average power is 103.51~mW in 10 seconds. The power consumption of the adaptive frequency hopping system BLE is 59.93~$\mu$Ah, which is 2.66~$\mu$Ah more than the non-adaptive system, an increase of only 4.6\%. The average power is 108.18~mW, which is 5~mW higher than the non-adaptive system, an increase of 4.8\%. The additional power consumption of the comprehensive real-time channel monitoring is about 4.6--4.8\%.

\subsection{Authentication Scheme Analysis}

\subsubsection{Authentication protocol security Analysis}

This section analyzes the security of the proposed scheme against common attacks to evaluate its effectiveness in addressing security threats and ensuring communication security.

In a impersonation  attack, an attacker impersonates a legitimate user or device to gain unauthorized access or perform malicious operations. In network communications, a impersonation  attack may involve falsifying identities, fake authentication, or tricking systems into performing unauthorized operations. For example, an attacker may disguise themselves as a legitimate user and send fake authentication information to the system to gain access to system resources.

A replay attack occurs when an attacker illegally intercepts a valid data transmission during network communication and then re-transmits the data to the receiver at a later time in order to deceive the receiver. This attack method can be used to bypass authentication systems and perform unauthorized operations.

The following is the security analysis against impersonation attack and replayattack, where impersonation attack is divided into two types.

\paragraph{Impersonation attack 1}

\textbf{Attack Model: }
In this attack, the attacker eavesdrops on the broadcast content, extracts the data from the broadcast, inserts it into their own broadcast, and then transmits it in an attempt to establish a connection with the vehicle.

\textbf{Lemma 1: }The whitelisting mechanism is able to defend against broadcast content injection attacks by utilizing a whitelist of radio frequency addresses on the vehicle. Since the attacker's broadcast address is not included in the whitelist, the connection attempt by the attacker fails.

\paragraph{Impersonation attack 2}

In this attack, due to the loss or theft of the key fob, the attacker can establish a communication connection with the vehicle using the key fob. The certificate for the lost key fob, $fob1.crt$, is revoked. We divide this into two cases:

1) Only the certificate $fob1.crt$ for the lost key fob is revoked, and the authentication system is not updated, i.e., no new key pseudo-ID ($PIDfob2$), public-private key pair ($fob2.pk$, $fob2.sk$), certificate ($fob2.crt$), and initial symmetric key ($rand\_init$) are generated for the new key.

2) For the new key, a new key pseudo-ID ($PIDfob2$), public-secret key pair ($fob2.pk$, $fob2.sk$), certificate ($fob2.crt$), and initial symmetric key ($rand\_{init}$) are generated (since the public key certificate is public, the attacker $S_2$ will install the new public certificate $fob2.crt$).

The following discusses the defense of the proposed scheme against this attack according to the above two cases:

\begin{figure}[h]
    \centering
    \includegraphics[width=0.48\textwidth]{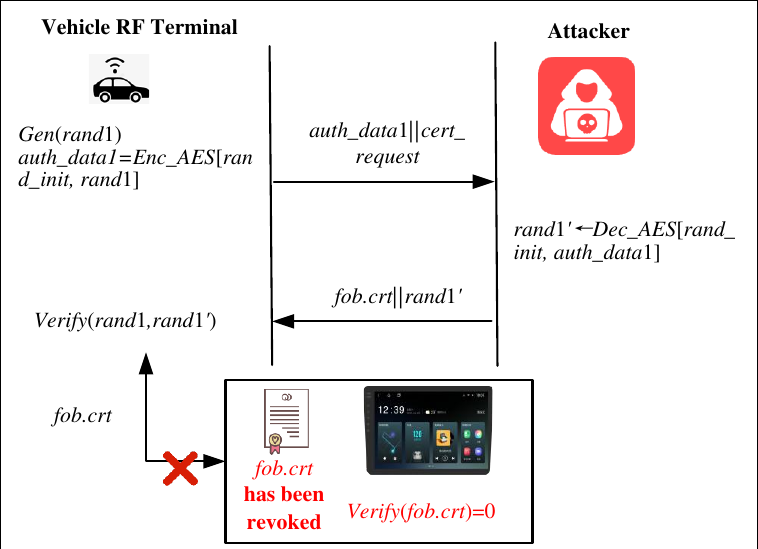}
    \caption{Authentication of invalid key certificate failed}
    \label{fig:attack1}
\end{figure} 
Attack Model: 
In this attack, the attacker exploits a revoked certificate to perform an impersonation attack, as shown in Figure \ref{fig:attack1}. The attacker uses a stolen key to establish a connection with the vehicle. The Vehicle RF Terminal proceeds with the authentication process by generating a random number \( rand1 \), encrypting it using the AES algorithm with the \( rand\_init \) encryption key, and obtaining the cipher text \( auth\_data1 \). The attacker then decrypts \( auth\_data1 \) to retrieve \( rand1' \) and sends the key \( fob.crt \) along with \( rand1' \). The Vehicle RF Terminal compares \( rand1' \) with the originally generated \( rand1 \) and finds them identical. It then forwards \( fob.crt \) to the vehicle’s certificate authentication system for validation.  

Lemma 2: The proposed revoked certificate verification scheme is able to defend against impersonation attacks by detecting revoked certificates during the authentication process. When the vehicle’s certificate authentication system verifies the key \( fob.crt \), it identifies the certificate as revoked, leading to authentication failure. Consequently, the system successfully prevents the attacker's impersonation attempt.

\begin{figure}[h]
    \centering
    \includegraphics[width=0.48\textwidth]{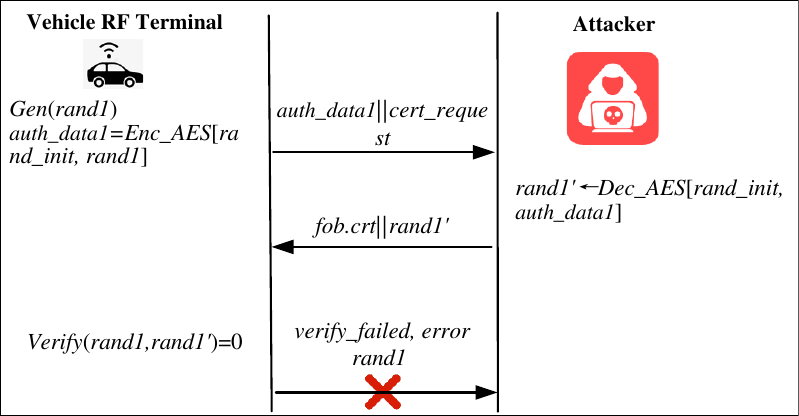}
    \caption{Authentication of invalid key certificate failed}
    \label{fig:attack2}
\end{figure} 
\textbf{Attack Model:  }
In this attack, the attacker attempts a counterfeit attack on the certification system, as shown in Figure \ref{fig:attack2}. The Vehicle RF Terminal generates a random number \( rand1 \) and encrypts it using the AES algorithm. Since the certificate authentication system has been updated, the encryption key has also been updated to \( rand\_{init2} \), resulting in the cipher text \( auth\_data1 = Enc\_AES[rand\_{init2}, rand1] \). However, the attacker only possesses the original initial symmetric key \( rand\_{init} \) and attempts to decrypt \( auth\_data1 \) using the AES algorithm in Equation \eqref{eq:attack}.

\textbf{Lemma 3:  }
The proposed updated certification system is able to defend against counterfeit attacks by periodically updating encryption keys. Since the attacker only holds the outdated key \( rand\_{init} \), they cannot correctly decrypt the cipher text \( auth\_data1 \), preventing unauthorized access to the vehicle system.  

\begin{equation}
     rand1' = Dec\_AES [rand\_init, auth\_data1]
    \label{eq:attack}
\end{equation}

Since $rand\_init1$ and $rand\_init2$ are not equal, the correct $rand1$ cannot be decrypted. The attacker decrypts the random number $rand1'$, sends $fob2.crt$ and $rand1'$, and the vehicle RF terminal determines that $rand1'$ is not equal to the random number $rand1$ generated in the previous step ($Verify(rand1', rand1) = 0$). Then the authentication fails, and the system successfully defends against the counterfeiting attack by attacker. With traditional RKE, if the key is lost or stolen, the vehicle can still be controlled. However, as we have seen in the analysis of the proposed scheme's defense against impersonation attacks, if the key is lost or stolen, the vehicle can be prevented from being controlled by the lost or stolen key by promptly revoking the certificate or updating the certificate system. This further improves the security compared to the traditional RKE system.

\paragraph{Replay attack}

A replay attack occurs when an attacker illegally intercepts a valid data transmission during network communication and then re-transmits the data to the receiver at a later time in order to deceive the receiver. This attack method can be used to bypass authentication systems and perform unauthorized operations.

\textbf{Attack Model: } 
In this attack, the attacker records the communication between the legitimate key fob and the vehicle by capturing and replaying the messages. The replayed messages contain control commands intended to illegally unlock the vehicle door.  

\begin{figure}[h]
    \centering
    \includegraphics[width=0.48\textwidth]{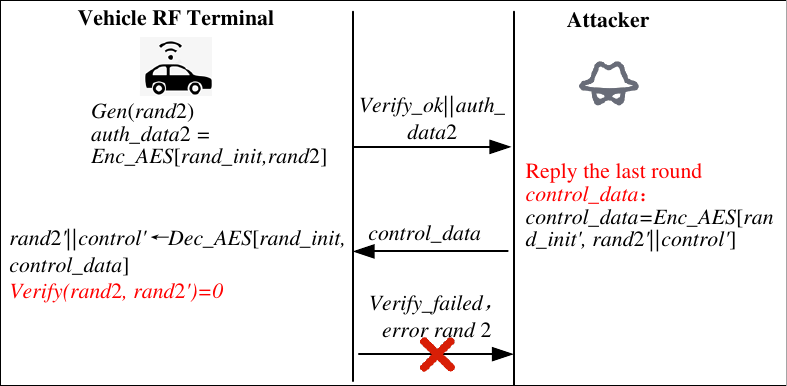}
    \caption{ Authentication of invalid key certificate failed}
    \label{fig:attack3}
\end{figure}

\textbf{Lemma 4: } 
The proposed replay attack defense scheme is able to defend against replay attacks by updating the symmetric encryption key in each authentication round. As shown in Figure \ref{fig:attack3}, the attacker replays the recorded encrypted control command packet \( control\_data = Enc\_AES[rand\_init', rand2 || control] \). The vehicle RF terminal attempts to decrypt \( control\_data \) using \( rand\_{init} \) to obtain \( rand2 || control = Dec\_AES[rand\_{init}, control\_data] \). However, since the symmetric encryption key was updated in the previous round (from \( rand\_init' \) to \( rand\_init \)), the vehicle RF terminal cannot correctly decrypt \( rand2 \), resulting in \( Verify(rand2', rand2) = 0 \). Consequently, the command is not executed, effectively preventing the replay attack.

\begin{table*}[htbp]
\centering
\caption{Comparison of Overhead: Un-deployed vs. Deployed Authentication Systems}
\label{tab:overhead_comparison}
\begin{tabular}{|c|c|c|c|c|}
\toprule
Memory Region & \multicolumn{2}{c|}{Un-deployed} & \multicolumn{2}{c|}{Deployed} \\
\cline{2-5}
 & Used Size & Usage Percentage & Used Size & Usage Percentage \\
\midrule
FLASH & 236308 B & 22.54\% & 289152 B & 27.58\% \\
RAM & 39916 B & 15.23\% & 44376 B & 16.93\% \\
IDT\_LIST & 0 GB & 0.00\% & 0 GB & 0.00\% \\
\bottomrule
\end{tabular}
\end{table*}
\subsubsection{Authentication protocol overhead Analysis}

The AES symmetric encryption method used in this solution provides high security while requiring a key length that is much smaller than asymmetric encryption algorithms. For example, at the same level of security, AES only requires a 128-bit key, while RSA requires 3072 bits and ECC requires 256-383 bits. This efficient key length makes AES excel in environments with limited computing resources, reducing storage and computing overheads and truly embodying the advantages of lightweight encryption. 
The certificate and AES encryption authentication protocol based on ECDSA proposed in this paper only requires one certificate authentication on the in-vehicle terminal and five symmetric encryptions between the vehicle RF terminal and the car key. Compared to RSA-based certificate authentication schemes, this approach saves key and certificate storage space. The proposed initial symmetric key update mechanism ensures that the symmetric key for each communication round is different. Compared to the Diffie-Hellman key exchange mechanism, this method saves storage space and key exchange time. Table \ref{tab:overhead_comparison} compares the memory and data storage overheads incurred by the key terminal before and after deploying the authentication system. From the calculations, it can be observed that deploying the authentication system results in only a 22.36\% increase in storage overhead and an 11.16\% increase in memory overhead, demonstrating its efficiency in resource utilization.

\section{Summary}
This paper proposes an adaptive RKE secure communication system to address the security vulnerabilities of traditional RKE systems, which suffer from fixed-frequency transmission and lack of authentication mechanisms. By introducing a real-time adaptive frequency-hopping scheme and a lightweight authentication mechanism based on ECDSA and AES, the system enhances both communication reliability and security. A prototype system was developed and tested, demonstrating significant improvements in stability and resistance to security threats. The results confirm the feasibility of the proposed approach, laying a foundation for future advancements in intelligent vehicle security.

Future work will focus on optimizing key management, exploring alternative communication protocols, and integrating hardware security modules (HSM) to enhance efficiency and robustness.

\newpage

\vfill

\end{document}